\documentclass[aps,prd,floatfix,superscriptaddress,balancelastpage,nofootinbib,amsmath,showpacs,twocolumn]{revtex4-1}
\usepackage{graphicx,epstopdf,epsfig,bm,dcolumn,natbib,amsmath,amssymb,multirow}

\usepackage{color}         

\usepackage[
	colorlinks=true,
	citecolor=black,
	linkcolor=black,
	urlcolor=black,
	hypertexnames=false]{hyperref}

\newcommand{\beq}{\begin{equation}} \newcommand{\eeq}{\end{equation}}
\newcommand{\bea}{\begin{eqnarray}} \newcommand{\eea}{\end{eqnarray}}

\def\lsim{\mathrel{\raise.3ex\hbox{$<$\kern-.75em\lower1ex\hbox{$\sim$}}}}
\def\gsim{\mathrel{\raise.3ex\hbox{$>$\kern-.75em\lower1ex\hbox{$\sim$}}}}

\begin{document}

\title{A 3.55 keV Line from Exciting Dark Matter without a Hidden Sector}

\author{Asher Berlin}
\affiliation{Enrico Fermi Institute, University of Chicago, Chicago, IL 60637}
\affiliation{Kavli Institute for Cosmological Physics, University of Chicago, Chicago, IL 60637}
\author{Anthony DiFranzo}
\affiliation{Center for Particle Astrophysics, Fermi National Accelerator Laboratory, Batavia, IL 60510}
\affiliation{Department of Physics and Astronomy, University of California, Irvine, CA 92697}
\author{Dan Hooper}
\affiliation{Center for Particle Astrophysics, Fermi National Accelerator Laboratory, Batavia, IL 60510}
\affiliation{Department of Astronomy and Astrophysics, University of Chicago, Chicago, IL 60637}

\begin{abstract}

Models in which dark matter particles can scatter into a slightly heavier state which promptly decays to the lighter state and a photon (known as eXciting Dark Matter, or XDM) have been shown to be capable of generating the 3.55 keV line observed from galaxy clusters, while suppressing the flux of such a line from smaller halos, including dwarf galaxies. In most of the XDM models discussed in the literature, this up-scattering is mediated by a new light particle, and dark matter annihilations proceed into pairs of this same light state. In these models, the dark matter and mediator effectively reside within a hidden sector, without sizable couplings to the Standard Model. In this paper, we explore a model of XDM that does not include a hidden sector. Instead, the dark matter both up-scatters and annihilates through the near resonant exchange of a $\mathcal{O}(10^2)$ GeV pseudoscalar with large Yukawa couplings to the dark matter and smaller, but non-neglibile, couplings to Standard Model fermions. The dark matter and the mediator are each mixtures of Standard Model singlets and $SU(2)_W$ doublets. We identify parameter space in which this model can simultaneously generate the 3.55 keV line and the gamma-ray excess observed from the Galactic Center, without conflicting with constraints from colliders, direct detection experiments, or observations of dwarf galaxies.

\end{abstract}

\pacs{95.35.+d, 95.85.Pw; FERMILAB-PUB-15-009-A}

\maketitle

\section{Introduction}
The nature of dark matter remains one of the most elusive and longstanding problems in physics today. As a consequence, much attention has been given to observational anomalies that can be plausibly interpreted in terms of dark matter interactions. One such signal is an approximately 3.55 keV X-ray line that has been observed from a number of galaxy clusters, as well as from the nearby Andromeda Galaxy.

The first reported evidence for the 3.55 keV line was found in data from the XMM-Newton satellite, from the directions of a stacked sample of 73 low redshift galaxy clusters \cite{Bulbul:2014sua}. Shortly thereafter, a similar line was reported from the directions of the Perseus Cluster and the Andromeda Galaxy \cite{Boyarsky:2014jta}. A study of XMM-Newton data also suggests the existence of a 3.55 keV line from the direction of the Milky Way's center \cite{Boyarsky:2014ska} (see also, however, Ref~\cite{Riemer-Sorensen:2014yda}). More recently, the line was identified within Suzaku data from the Perseus Cluster \cite{Urban:2014yda}. 

A number of interpretations for these observations have been proposed. On the one hand, it has been suggested that atomic transitions (such as those associated with the chlorine or potassium ions, Cl-XVII and K-XVIII, for example~\cite{Jeltema:2014qfa}) might be responsible for the line, although the viability of this explanation is currently unclear \cite{Bulbul:2014ala, Boyarsky:2014paa, Jeltema:2014mla}. Alternatively, decaying dark matter particles could generate such an X-ray line. Particularly well motivated is dark matter in the form of an approximately 7 keV sterile neutrino, which decays through a loop to a photon and an active neutrino. If one assumes that all of the dark matter consists of 7 keV sterile neutrinos, the observed X-ray line flux implies a mixing angle of $\sin^2(2\theta) \sim 7 \times 10^{-11}$. With such a small degree of mixing, however, the standard Dodelson-Widrow mechanism of production via the collision-dominated oscillation conversion of thermal active neutrinos~\cite{Dodelson:1993je} leads to an abundance of sterile neutrinos that corresponds to only a few percent of the total dark matter density, thus requiring additional resonant or otherwise enhanced production mechanisms. Alternatively, sterile neutrinos with a larger mixing angle of $\sin^2(2\theta) \sim 3 \times 10^{-10}$ could naturally constitute roughly 10\% of the dark matter abundance, and decay at a rate that is sufficient to generate the observed line flux.

Interpretations of the X-ray line in terms of decaying dark matter are in considerable tension, however, with studies of galaxies using Chandra and XMM-Newton data \cite{Anderson:2014tza} and dwarf spheroidal galaxies using XMM-Newton data \cite{Malyshev:2014xqa}, which do not detect a line at the level predicted by decaying dark matter scenarios.  One way to potentially reconcile the intensity of the line observed from clusters with the null results from dwarfs and other smaller systems is to consider the class of scenarios known as eXciting Dark Matter (XDM) \cite{Finkbeiner:2007kk, Pospelov:2007xh, Finkbeiner:2014sja, Frandsen:2014lfa}. In such models, the collisions of dark matter particles can cause them to up-scatter into an excited state, $\chi_1 \chi_1 \rightarrow \chi_2 \chi_2$ or $\chi_1 \chi_1 \rightarrow \chi_1 \chi_2$. For a mass splitting of $m_{\chi_2}-m_{\chi_1} \simeq \, 3.55$ keV, the subsequent decays of the slightly heavier state can generate a 3.55 keV photon, $\chi_2 \rightarrow \chi_1 \gamma$. Critical to the problem at hand are the kinematics of the XDM scenario, which introduce a velocity threshold for up-scattering, suppressing the X-ray flux from dwarf galaxies (and, to a lesser extent, from larger galaxies)\cite{Cline:2014vsa,Finkbeiner:2014sja}. Within the paradigm of XDM, the observations of clusters, galaxies, and dwarf galaxies can be mutually consistent for dark matter masses between approximately 40 GeV and 10 TeV \cite{Cline:2014vsa}, covering the mass range generally associated with conventional WIMPs.

If up-scattering WIMPs are responsible for the 3.55 keV line, one might also imagine that the same dark matter species could generate the excess of GeV-scale gamma-rays observed from the region surrounding the Galactic Center~\cite{Goodenough:2009gk, Hooper:2010mq, Hooper:2011ti, Abazajian:2012pn, Hooper:2013rwa, Gordon:2013vta, Macias:2013vya, Abazajian:2014fta, Daylan:2014rsa,Calore:2014xka}. This signal, identified within data from the Fermi Gamma-Ray Space Telescope, exhibits a spectrum and morphology that are in good agreement with that anticipated from dark matter annihilations. This data has been explored by several groups independently, including recently the Fermi Collaboration~\cite{fermigc}. Assuming annihilations to $b\bar{b}$, for example, dark matter particles with a mass of $m_{\chi}\sim$ 35-65 GeV and a cross section of $\langle\sigma v\rangle \sim10^{-26}$ cm$^3/$s provide a good fit to the observed excess \cite{Calore:2014nla}.

The primary challenge in developing a viable XDM model for the 3.55 keV line is that the up-scattering rate must be very high, several orders of magnitude larger than the annihilation rate. One way to realize this is to consider dark matter that scatters through a light mediator and annihilates into pairs of the same mediator. This naturally leads to an up-scattering rate that is enhanced by a factor of $\sim (m_{\chi}/M_{\rm Med})^4$ relative to the annihilation rate. As this phenomenology can be realized without the dark matter or mediator possessing any sizable couplings to the Standard Model (SM), these scenarios are sometimes called ``hidden sector'' models. Examples of such proposals include models with a massive vector (hidden photon) or a massive scalar (hidden Higgs) that couples directly to the dark matter, but interacts with the SM only through a very small degree of kinetic or mass mixing. As a result, the dark sector and SM are effectively sequestered from one another. As this class of possibilities has been explored previously in some detail~\cite{Finkbeiner:2014sja,Frandsen:2014lfa,Cline:2014kaa,Cheung:2014tha,Cline:2014eaa}, we do not consider it here. Instead, we explore models in which the dark matter annihilates directly into SM fermions (for an earlier investigation in this direction, see Ref.~\cite{Okada:2014zea}). By introducing a resonant mediator with a hierarchy of couplings ($g_{\rm dark} >> g_{\rm SM}$), it is possible to accomplish similar phenomenology without a light mediator. We identify such a model that can simultaneously explain the 3.55 keV line from Galaxy Clusters and the Galactic Center gamma-ray excess. We find viable parameter space in our model that is consistent with all current collider, direct detection, and indirect detection constraints.

\section{The Kinematics of eXciting Dark Matter}
\label{kinematics}
If the 3.55 keV signal is due to dark matter, the model responsible needs to address why this signal is not seen from dwarf galaxies (and, to a lesser extent, from larger galaxies). As the up-scattering rate in the XDM scenario depends strongly on the dark matter velocity dispersion in such systems, this framework provides a simple mechanism to suppress the line flux predicted from smaller halos. 

The velocity averaged cross section for up-scattering is given by:
\beq
\langle\sigma v\rangle = \sigma_0 v_t \gamma, 
\eeq
where the normalization, $\sigma_0$, is taken to be a free parameter and $\gamma$ accounts for the effect of the threshold velocity on the up-scattering rate:
\beq
\gamma = \left\langle\sqrt{v^2/v_t^2-1} \,\,\, \Theta\left(v-v_t\right) \right\rangle.
\eeq 
The quantity $v_t$ is the threshold velocity, given by:
\beq
v_t = 2 \sqrt{N\frac{\delta m_{\chi}}{m_{\chi}}},
\label{vt}
\eeq
where $\delta m_{\chi}$ (taken to be $\simeq 3.55$ keV) is the mass splitting between $\chi_2$ and $\chi_1$ and $N=1$ (2) for up-scattering to $\chi_1 \chi_2$ ($\chi_2 \chi_2$). 

In the limit of $\delta m_{\chi} =0$ (and $v_t=0$), the standard $\langle\sigma v\rangle = \sigma_0 v$ is recovered. For larger mass splittings, however, the up-scattering rate and corresponding line flux will be suppressed in smaller systems, where typical velocities are lower.

To obtain up-scattering rates in dwarfs, galaxies, and clusters that are each compatible with the reported observations,  Ref.~\cite{Cline:2014vsa} finds that a threshold velocity of $v_t \simeq 20-245$ km/s is required (at the $\delta\chi^2 < 3$ level, and assuming that the excited state decays promptly). Combining this with Eq.~\ref{vt} (where $N=2$), this implies $m_{\chi} \sim $\,\,40 GeV-10 TeV.  Annihilating dark matter particles near the low end of this mass range are also well suited to account for the Galactic Center gamma-ray excess.

\section{Model Building}
\label{model}

\subsection{Up-scattering}
\label{upscattering}

There are two classes of scenarios in which an excited state could be presently decaying in order to generate the observed 3.55 keV line. First, if the excited state has a lifetime on the order of the age of the Universe or longer, a population of such particles could have been produced in the early universe. Primordial excitations, however, do not lead to a relative suppression in dwarf galaxies, and thus suffer from the same challenges in explaining the 3.55 keV line as ordinary decaying dark matter. Alternatively, if the excited state is short lived (millions of years or less) collisions between dark matter particles must lead to an up-scattering rate that is sufficient to perpetually populate these excitations in galaxy clusters. It is this second case that we consider here.

In order for XDM to generate the flux of 3.55 keV photons observed from galaxy clusters, very large cross sections for up-scattering are required, in the approximate range of $\sigma v \left( \chi_1 \chi_1 \to \chi_2 \chi_2 \right) \sim \left(m_\chi/50 \text{ GeV} \right)^2 \times 10^{-18} \text{ cm}^3 /\text{s}$. 
In addition to being very large in and of itself, this value for the up-scattering cross section is several orders of magnitude larger than the annihilation cross section needed to generate the Galactic Center gamma-ray excess, or to obtain a thermal relic abundance in agreement with the measured dark matter density.

\begin{figure}
\includegraphics[width=3.0in]{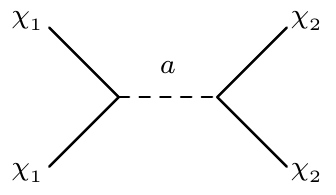} 
\caption{The dominant Feynman diagram for up-scattering in our model.}
\label{upscatterdiagram}
\end{figure}

In light of this, it is interesting to consider the upper limit imposed on dark matter scattering from the point of view of perturbativity and unitarity. In this paper, we will focus on up-scattering through a resonant $s$-channel pseudoscalar, $a$ (see Fig.~\ref{upscatterdiagram}). We will further assume that the dark matter and its excited state, $\chi_{1,2}$, are each Majorana fermions with nearly degenerate masses, $m_{\chi_1} \approx m_{\chi_2} \equiv m_\chi$ (collectively constituting a pseudo-Dirac fermion). The scalar ($J=0$) bilinear involved in this interaction, $\bar{\chi}i \gamma^5 \chi$, being even under charge, $\mathcal{C}=\left( -1\right)^{L+S}$, and odd under parity, $\mathcal{P} = \left( -1 \right)^{L+1}$, implies that this operator only acts on incoming dark matter pairs with zero spin and orbital angular momentum, $J=S=L=0$. As a result, scattering through an $s$-channel pseudoscalar is purely $s$-wave, and the unitarity bound (see {\it e.g}. Ref.~\cite{Griest:1989wd}) on up-scattering is given by (assuming $\chi_2$ is self-conjugate):
\begin{equation}
\label{eq: unitarity1}
\sigma v \leq \frac{2 \pi}{m_{\chi}^2 v} \approx \left( \frac{50 \text{ GeV}}{m_{\chi}}\right)^2 \left( \frac{0.003}{v} \right) \times 10^{-17} \text{ cm}^3/\text{s}
~,
\end{equation}
where $v \sim 0.003$ is the typical dark matter relative velocity in a galaxy cluster. For comparison, note that this upper limit is much stronger than that derived from self-scattering in objects such as the Bullet Cluster, $\sigma v \lesssim \left( \frac{m_\chi}{50 \text{ GeV}} \right) \left( \frac{v}{0.003} \right) \times 10^{-14} \text{ cm}^3/\text{s}$~\cite{Weinberg:2013aya}. 

A more explicit bound on the couplings of the theory arises if one parametrizes the Lagrangian responsible for up-scattering as follows:
\begin{equation}
\label{eq: simplemodel}
\mathcal{L} \supset \lambda^a_{11} a  \bar{\chi}_1 i \gamma^5 \chi_1 + \lambda^a_{22} a  \bar{\chi}_2 i \gamma^5 \chi_2 + \lambda^a_{12} a  \bar{\chi}_1 i \gamma^5 \chi_2.
\end{equation}

If we assume that $\chi_1$ and $\chi_2$ couple to the pseudoscalar, $a$, with approximately equal strength, we can use Eq.~\ref{eq: unitarity1} to deduce:
\begin{equation}
\label{eq: unitarity2}
\lambda^a_{11,22} \leq \sqrt{2\pi} \left[ \frac{\left( s - m_a^2\right)^2 + m_a^2 \Gamma_a^2}{s \left(s-4m_{\chi}^2\right)} \right]^{1/4}
~,
\end{equation}
where $m_a$ and $\Gamma_a$ are the mass and width of $a$, respectively. Therefore, perturbative unitarity of the theory in the non-relativistic and ultra-relativistic regimes requires: 
\begin{equation}
\label{eq: unitarity3}
\lambda^a_{11,22} \lesssim \begin{cases}20 \times \left( \delta / 0.1 \right)^{1/2} \left( 0.003 / v \right)^{1/2} , & \sqrt{s} \approx 2m_{\chi}  \\ 2.5, & \sqrt{s} \sim \infty, \end{cases}
\end{equation}
where $\delta \equiv |1- \left( m_a/2 m_{\chi}\right)^2 | \ll 1$. In the first line of Eq.~\ref{eq: unitarity3}, we have assumed that the width $\Gamma_a$ is sufficiently small such that $\Gamma_a/m_a \ll \delta$. We will show later that these conditions will be satisfied within the most viable parameter space for generating a large cross section for up-scattering. We see that in the non-relativistic regime, the upper limit on $\lambda^a_{11,22}$ from perturbative unitarity is weaker than one generically expects from perturbativity of the theory, $\lambda^a_{11,22} \lesssim 4 \pi$. Throughout our analysis, we will consider Yukawa couplings as large as $\lambda^a_{11,22} \sim 5$ in the non-relativistic regime. If $\chi_{1,2}$ only couple to $a$, then large values of $\lambda^a_{11,22}$ will contribute positively to its beta function and cause $\lambda^a_{11,22}$ to grow rapidly at higher energies. By considering such large values of this coupling, we implicitly require that new physics (such as couplings to new gauge bosons) come in at higher energies in order to stabilize $\lambda^a_{11,22} \lsim 2.5$ in the high-energy limit.

In the low velocity limit, the cross section for $\chi_1 \chi_1$ to become excited to $\chi_2 \chi_2$ is given by:
\begin{eqnarray}
\sigma v (\chi_1 \chi_1 \rightarrow \chi_2 \chi_2) &\approx& (v^2 - v^2_{t})^{1/2}  \\
&\times& \frac{ 2 m^2_{\chi} [ \lambda^a_{11} \lambda^a_{22}]^2 }{\pi     [ ( 4m^2_{\chi}-m^2_a)^2 +m^2_a \Gamma^2_a]},  \nonumber 
\end{eqnarray}
where $v_t$ is as defined in Eq.~\ref{vt}. The up-scattering of $\chi_1 \chi_1$ into $\chi_1 \chi_2$ is subdominant, but for a somewhat subtle reason. As we will see later in this paper, after diagonalizing into mass eigenstates, one of the fields $\chi_1$ or $\chi_2$ generally has a mass term with the ``wrong sign'', requiring a transformation $\chi \rightarrow i \gamma^5 \chi$. This leaves the $a\bar{\chi}_1 i \gamma^5 \chi_1$ and $a\bar{\chi}_2 i \gamma^5 \chi_2$ interaction terms as written in Eq.~\ref{eq: simplemodel}, but changes the mixed term into $a \bar{\chi}_1 \chi_2$, resulting in the suppression of the corresponding up-scattering cross section by an additional factor of $(v^2 - v^2_{t})$.

\subsection{Annihilation and Coannihilation}
\label{annihilation}

In the previous subsection, we showed that the very large up-scattering rates required for the 3.55 keV line can be generated through the resonant exchange of a pseudoscalar, $a$, but at the cost of introducing $\mathcal{O}(1)$ Yukawa couplings into the dark matter sector. If $\chi_1$ is to be populated thermally in the early universe, however, it must also have non-zero couplings to the SM. Furthermore, if the couplings of $a$ to the SM are comparable to its couplings to the dark matter, then the annihilation cross section during freeze-out will be many orders of magnitude larger than that needed to generate a thermal relic abundance of $\chi_1$ consistent with $\Omega_{\chi} h^2 \sim 0.12$. Instead, there must be a large hierarchy between the couplings of $a$ with the dark matter and with the SM.

One way to generate a very small coupling between the $a$ and SM fermions is through mass-mixing with the heavy pseudoscalar in a two-Higgs doublet model (2HDM). Such a scenario has been discussed previously within the context of the Galactic Center gamma-ray excess~\cite{Ipek:2014gua}. The idea is to introduce a scalar potential involving a parity-odd singlet pseudoscalar, $a_0$, along with a second Higgs doublet in the framework of a Type-II 2HDM. The two Higgs doublets, each with hypercharge of $+1/2$, are denoted as $H_{d,u}$, and the corresponding pseudoscalar of the 2HDM sector is written as $A_0$. After $a_0$ and $A_0$ mix, the light and heavy mass eigenstates of the CP-odd sector will be written as $a$ and $A$, respectively.

\begin{figure}
\includegraphics[width=3.0in]{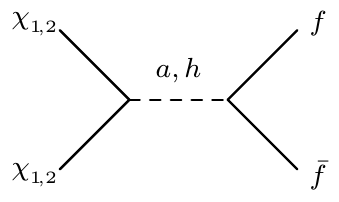} 
\caption{The dominant Feynman diagram for the annihilation or coannihilation of $\chi_1 \chi_1$, $\chi_1 \chi_2$, or $\chi_2 \chi_2$.}
\label{annihdiagram}
\end{figure}

The Higgs portal between the dark matter and the SM emerges from the trilinear interaction of the scalar potential involving $a_0$, $H_d$, and $H_u$. More specifically, the terms of the scalar potential relevant for the annihilation and coannihiation of $\chi_{1,2}$ are given by:
\begin{equation}
\label{eq: potential}
V_\text{scalar} \supset V_\text{2HDM} + \frac{1}{2} m_{a_0}^2 a_0^2 + \left( i B_a a_0 H_d^\dagger H_u + \text{h.c.} \right),
\end{equation}
where $V_\text{2HDM}$ is the most general CP-conserving 2HDM potential corresponding to a Type-II 2HDM, and $B_a$ is a dimensionful parameter governing the strength of mixing in the Higgs portal. In order to suppress flavor changing neutral currents at tree-level, Type-II 2HDMs involve a $\mathbb{Z}_2$ symmetry under which $H_d \to -H_d$ and $H_u \to H_u$. We have assumed that this symmetry is softly broken by dimensionful couplings, such as by the Higgs portal interaction involving $B_a$ in Eq.~\ref{eq: potential}, and similar terms within the 2HDM scalar potential. For simplicity, we assume that CP is conserved in the full potential of Eq.~\ref{eq: potential}, which implies that $B_a$ is real and that $a_0$ and $A_0$ do not develop vacuum expectation values. Once electroweak symmetry breaking is induced by $\langle H_{d,u} \rangle = v_{d,u} / \sqrt{2}$, $H_d$ and $H_u$ can be written in terms of the scalar mass eigenstates of the 2HDM potential:
\begin{align}
H_d &= \frac{1}{\sqrt{2}} \begin{pmatrix} -\sqrt{2}~s_\beta H^+ + \sqrt{2}~c_\beta G^+\\ v_d -s_\alpha h + c_\alpha H - i s_\beta A_0 + i c_\beta G  \end{pmatrix}
~,
\nonumber \\
H_u &= \frac{1}{\sqrt{2}} \begin{pmatrix} \sqrt{2}~c_\beta H^+ + \sqrt{2}~s_\beta G^+\\ v_d  + c_\alpha h + s_\alpha H + i c_\beta A_0 + i s_\beta G  \end{pmatrix}
~,
\end{align} 
where $h$, $H$ are the light and heavy CP-even Higgs bosons, $H^\pm$ the charged Higgs, $G$ and $G^\pm$ the neutral and charged Goldstones bosons, and $A_0$ is the pseudoscalar of the 2HDM sector. $c_\beta$ and $s_\beta$ are the cosine and sine of $\beta$, defined by $\tan{\beta} \equiv v_u / v_d$ and $\sqrt{v_d^2 + v_u^2} = v = 246$ GeV, and $c_\alpha$ and $s_\alpha$ are the cosine and sine of the mass mixing angle of the CP-even scalars, $\alpha$. We will remove all dependence on $\alpha$ by choosing to work in the alignment limit throughout, where $\sin({\beta-\alpha})=1$ and the $h$ couplings are SM-like. We will further assume that the masses of $A_0$, $H$, and $H^\pm$ are decoupled, with values at a scale around 1 TeV.

Mass mixing between the pseudoscalars $a_0$ and $A_0$ is induced by the coupling $B_a$ in Eq.~\ref{eq: potential}. The mass-squared matrix of the CP-odd sector in the $(a_0, A_0)$ basis is written as:
\begin{equation}
M^2_\text{CP-odd} = \begin{pmatrix} m_{a_0}^2 & - B_a v \\ -B_a v & m_{A_0}^2 \end{pmatrix}
~,
\end{equation}
where $m_{A_0}$ is the mass of the the pseudoscalar $A_0$ in the 2HDM potential. Diagonalizing $M^2_\text{CP-odd}$ leads to the mass eigenstates $a$ and $A$ such that:
\begin{align}
\begin{pmatrix} a_0 \\ A_0 \end{pmatrix} &= \begin{pmatrix} \cos{\theta} & \sin{\theta} \\ -\sin{\theta} & \cos{\theta} \end{pmatrix} \begin{pmatrix} a \\ A \end{pmatrix},\\
m_{a,A}^2 &= \frac{1}{2} \left[m_{A_0}^2+m_{a_0}^2 \mp \sqrt{\left(m_{A_0}^2-m_{a_0}^2\right)^2 + 4 B_a^2 v^2}~\right],
\nonumber \\
\cos{\theta} &= \frac{1}{\sqrt{2}} \left( 1+ \frac{m_{A_0}^2-m_{a_0}^2}{\sqrt{\left( m_{A_0}^2 - m_{a_0}^2 \right)^2 + 4 B_a^2 v^2}} \right)^{1/2}. \nonumber
\end{align}

Throughout this work, we will consider values of $m_a \sim 100$ GeV, $m_A \sim m_H \sim m_{H^\pm} \sim 1$ TeV, and $\theta \sim  10^{-5}$. These choices of parameters uniquely determine $| B_a | \sim \mathcal{O}(0.1)$ GeV. Therefore, we will be working in the limit in which mixing is induced by small off-diagonal terms and $m_{a_0} \ll m_{A_0}$. As a result, the light pseudoscalar, $a$, is mostly singlet-like and the much heavier $A$ is mostly 2HDM-like.

Stringent constraints on new scalars and pseudoscalars can be derived from the results of searches for MSSM Higgs bosons at colliders~\cite{Chatrchyan:2013qga,Aad:2014vgg,Khachatryan:2014wca}. In particular, if $a$ has some sizable branching ratio to SM fermions, then the production of a $a$ in association with a $b$-jet can produce distinctive $bbb$ and $b \tau \tau$ events. In our case, however, the $a$ has suppressed couplings to quarks and leptons and very large couplings to dark matter, enabling collider searches for invisibly decaying light scalars and pseudoscalars to provide much stronger bounds~\cite{Zhou:2014dba,Lin:2013sca,Aad:2014vea}. Even these searches, however, yield extremely weak bounds for $\theta \lsim 10^{-3}$ since the production of $a$ depends on its suppressed couplings to SM quarks. In fact, for $m_a \sim 100$ GeV in the large $\tan{\beta}$ limit, the most stringent constraint comes from the contribution to $B_s \to \mu^+ \mu^-$, which results in the approximate upper bound $\theta \lesssim 0.1$~\cite{Ipek:2014gua}.


Due to the small mass splitting between $\chi_1$ and $\chi_2$, both of these states can play an important role in determining the thermal relic abundance of dark matter in this model.  Although we use the publicly available program micrOMEGAs~\cite{Belanger:2013oya} to calculate the relic abundance numerically, it is illustrative to consider analytic forms for the relevant cross sections in the low velocity limit (see Fig.~\ref{annihdiagram}):
\begin{eqnarray}
\label{anncs}
\sigma v (\chi_1 \chi_1 \rightarrow f \bar{f}) &\approx&  \frac{2 n_c}{\pi} \sqrt{1-\frac{m^2_f}{m^2_{\chi}}}  \bigg[\frac{\lambda_{11}^a \lambda^a_f m_{\chi}}{4m^2_{\chi}-m^2_a} \bigg]^2,  \\
\sigma v (\chi_2 \chi_2 \rightarrow f \bar{f}) &\approx&  \frac{2 n_c}{\pi} \sqrt{1-\frac{m^2_f}{m^2_{\chi}}}  \bigg[\frac{\lambda_{22}^a \lambda^a_f m_{\chi}}{4m^2_{\chi}-m^2_a} \bigg]^2, \nonumber \\
\sigma v (\chi_1 \chi_2 \rightarrow f \bar{f}) &\approx& \frac{n_c}{2\pi} \bigg(1-\frac{m^2_f}{m^2_{\chi}}\bigg)^{3/2}  \bigg[\frac{\lambda_{12}^h \lambda^h_f m_{\chi}}{4m^2_{\chi}-m^2_h}\bigg]^2, \nonumber
\end{eqnarray}
where $n_c=3 (1)$ for annihilation into quarks (leptons), and the widths of $a$ and $h$ should be included when near resonance. The first two of these processes are mediated by the exchange of the light pseudoscalar, $a$, which couples to the SM through mixing with the heavier pseudoscalar of the 2HDM,  $\lambda^a_f = -\sin \theta \, m_f \cot \beta/v$ for up-type fermions and $\lambda^a_f = -\sin \theta \, m_f \tan \beta/v$ for down-type fermions. The last of these processes is mediated by the SM-like scalar Higgs boson. For reasons that are similar to those described in the previous subsection for the process $\chi_1 \chi_1 \rightarrow \chi_1 \chi_2$, this process is $s$-wave and contributes in the low velocity limit. Here, $\lambda^h_{12}$ is the $\chi_1-\chi_2-h$ coupling (corresponding to the term $\lambda^h_{12} h  \bar{\chi}_1  \chi_2$ in the Lagrangian of Eq.~\ref{eq:OriginalLagrangian}), and $\lambda^h_f = -m_f/v$ is the coupling between the light scalar Higgs boson and SM fermions.

The process of thermal freeze-out in this model depends on the hierarchy of the annihilation and coannihilation cross sections described in Eq.~\ref{anncs}.  If $\sigma v (\chi_1 \chi_1 \rightarrow f \bar{f})$, $\sigma v (\chi_2 \chi_2 \rightarrow f \bar{f}) \gg \sigma v (\chi_1 \chi_2 \rightarrow f \bar{f})$, for example, each of the two species freeze-out largely independently of one another, followed by the decay $\chi_2 \rightarrow \chi_1 \gamma$, which increases the final abundance of the $\chi_1$ population. Alternatively, if $\sigma v (\chi_1 \chi_2 \rightarrow f \bar{f})$ is not negligible, these coannihilations will deplete the abundances of both species, and the total resulting dark matter abundance. The dark matter's annihilation cross section in the universe today can vary significantly depending on which of these processes dominates.  In the former case, we expect a comparatively large cross section, $\sigma v \sim (4-6) \times 10^{-26}$ cm$^3$/s, which is in tension with constraints from gamma-ray observations of dwarf spheroidal galaxies (especially in the case in which $2 m_\chi$ is near resonance, but slightly greater than $m_a$, for which the low velocity annihilation rate is further enhanced). If the cross section for $\chi_1 \chi_2$ coannihilations is substantial, however, the self-annihilation cross section required to generate the appropriate thermal relic abundance will be reduced,  allowing us to comfortably evade this constraint.


To summarize the major points of this section, if the $s$-channel exchange of the pseudoscalar, $a$, is to contribute to both the up-scattering and the annihilation of the dark matter, $a$ must have both large couplings to the dark matter and very small couplings to the SM. In our model,  these latter interactions arise from the mass-mixing of the $a$ with the 2HDM pseudoscalar, $A_0$, allowing the corresponding coupling to be highly suppressed.

\subsection{Decay and Mass Splitting}

\begin{table}[t]
\centering
\begin{tabular}{| c || c | c | c |}
\hline
Field & Charge & Spin \\ \hline \hline
$S_1$ & $(\mathbf{1}, \mathbf{1}, 0)$ & 1/2\\ \hline
$S_2$ & $(\mathbf{1}, \mathbf{1}, 0)$ & 1/2\\ \hline
$D_1$ & $(\mathbf{1}, \mathbf{2}, -1/2)$ & 1/2\\ \hline
$D_2$ & $(\mathbf{1}, \mathbf{2}, +1/2)$ & 1/2\\ \hline
$a_0$ & $(\mathbf{1}, \mathbf{1}, 0)$ & 0\\ \hline
$H_d$ & $(\mathbf{1}, \mathbf{2}, +1/2)$ & 0\\ \hline
$H_u$ & $(\mathbf{1}, \mathbf{2}, +1/2)$ & 0\\ \hline
\end{tabular}
\caption{The field content of the model described in this paper. The charges correspond to $SU(3)c \times SU(2)_W \times U(1)_Y$.}
\label{fields}
\end{table}

Up to this point, we have assumed that the dark matter and its excited state, $\chi_{1,2}$, are gauge singlets. While this is sufficient to obtain the desired rates for both up-scattering and annihilation, we must also require that the excited state, $\chi_2$, decays with a lifetime that is much shorter than the age of the universe. This requires the dark sector to couple to a charged state appearing in the loop-diagram responsible for the decay $\chi_2 \rightarrow \chi_1 \gamma$. 

A cosmologically short lifetime for $\chi_2$ can be easily accommodated by mixing a small $SU(2)_W$ doublet component into the dark sector, allowing $\chi_{1,2}$ to couple to the Higgs doublets, $H_{d,u}$, (and their associated $H^\pm$) as well as to $W^\pm$. Additionally, an active degree of freedom that is allowed to mix with $\chi_{1,2}$ generically introduces new charged states into the dark matter sector, analogous to charginos in the MSSM, that can enter into the loop-induced decay as well. The introduction of a small doublet component into the dark sector can also provide a natural explanation for the 3.55 keV mass splitting between the states $\chi_1$ and $\chi_2$.

In this regard, we follow closely the approach laid out in Ref.~\cite{Falkowski:2014sma}, wherein a vector-like pair of 2-component Weyl fermion SM gauge singlets, $S_1$ and $S_2$, and a vector-like pair of Weyl fermion $SU(2)_W$ doublets, $D_1$ and $D_2$, are introduced, the latter of which are assigned hypercharge $\mp 1/2$. Unlike in Ref.~\cite{Falkowski:2014sma}, however, which only considers long-lived primordial decays with mixing introduced by interactions with the SM Higgs doublet, we will consider interactions involving the two Higgs doublets and much shorter lifetimes for the excited state. 

In addition to the bare mass terms, in general, the singlet and doublet degrees of freedom can couple via Yukawa terms to one or both of the Higgs doublets, $H_{d,u}$. We will assume that the dark matter sector respects the $\mathbb{Z}_2$ symmetry of the 2HDM potential, which we enlarge to include $S_{1,2} \to - S_{1,2}$ (in addition to the usual $H_{d,u} \to \mp H_{d,u}$). As a result, the dark matter can only directly couple to $H_d$. We summarize our model's particle content in Table~\ref{fields}. The dark sector Lagrangian contains the following terms:
\begin{align}
\label{eq: lagrangian1}
- \mathcal{L} &\supset M_S S_1 S_2 + M_D D_1 D_2 + i y_S a_0 S_1 S_2 + i y_D a_0 D_1 D_2 
\nonumber \\
&~+ y_{11} S_1 D_1 H_d + y_{21} S_2 D_1 H_d 
\nonumber \\
&~+ y_{22} S_2 H^\dagger_d \cdot D_2 + y_{12} S_1 H^\dagger_d \cdot D_2 + \text{h.c.},
\end{align}
where 2-component Weyl and $SU(2)_W$ indices are implied. For simplicity, we take all of the couplings in Eq.~\ref{eq: lagrangian1} to be real and introduce an additional $\mathbb{Z}_2$ symmetry on the dark sector such that the lightest fermionic state is a stable dark matter candidate. For model building in a similar direction, see Ref.~\cite{preparation}.

\begin{figure}
\includegraphics[width=3.0in]{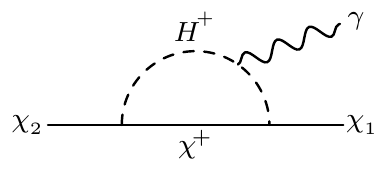} 
\includegraphics[width=3.0in]{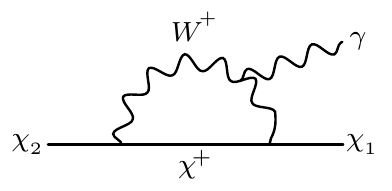} 
\caption{Feynman diagrams for the decay $\chi_2 \rightarrow \chi_1 \gamma$. Similar diagrams in which the photon is emitted off the $\chi^+$ also contribute.}
\label{decaydiagram}
\end{figure}

The dark matter $SU(2)_W$ doublets are parametrized as: 
\begin{equation}
D_1 = \begin{pmatrix} \nu_1 \\ E_1 \end{pmatrix} ~,~ D_2 = \begin{pmatrix} -E_2 \\ \nu_2 \end{pmatrix}
~,
\end{equation}
where $\nu_{1,2}$ and $E_{1,2}$ are the neutral and charged components of the doublets, respectively. $E_1$ and $E_2$ will mix to form an electrically charged Dirac fermion of mass $M_D$, which we will label as $\chi^{\pm}$. Since chargino searches at LEP generally exlude new charged fermions lighter than 100 GeV, we will take $M_D \gtrsim 100$ GeV. After elecroweak symmetry breaking, the neutral degrees of freedom, $S_{1,2}$ and $\nu_{1,2}$, will mix according to the following mass matrix (in the $S_1$-$S_2$-$\nu_1$-$\nu_2$ basis):
\begin{equation}
\label{eq: massmatrix}
M_0 = \begin{pmatrix} 0 & M_S & \frac{1}{\sqrt{2}}y_{11} v_d & \frac{1}{\sqrt{2}}y_{12} v_d \\  M_S & 0 & \frac{1}{\sqrt{2}}y_{21} v_d & \frac{1}{\sqrt{2}}y_{22} v_d \\ \frac{1}{\sqrt{2}}y_{11} v_d & \frac{1}{\sqrt{2}}y_{21} v_d & 0 & M_D \\ \frac{1}{\sqrt{2}}y_{12} v_d & \frac{1}{\sqrt{2}}y_{22} v_d & M_D & 0 \end{pmatrix}.
\end{equation}
The lightest mass eigenstate of $M_0$ will be the stable dark matter candidate, $\chi_1$, and the second lightest mass eigenstate the excited state, $\chi_2$. The gauge composition of $\chi_{1,2}$ can be written as:
\begin{equation}
\label{eq: mixingangles1}
\chi_{1,2} = N_{S_1}^{1,2} S_1 + N_{S_2}^{1,2} S_2 + N_{\nu_1}^{1,2} \nu_1 + N_{\nu_2}^{1,2} \nu_2
~.
\end{equation}

A large doublet component of $\chi_{1,2}$ is severely restricted by direct detection experiments because this usually introduces large couplings to the SM-like Higgs and to the $Z$. We therefore focus on the case in which $\chi_{1}$ and $\chi_2$ are largely singlet-like and $m_{\chi_1}\approx m_{\chi_2} \approx M_S \ll M_D$. In this small mixing limit, it is possible to derive approximate forms for the mixing angles:
\begin{align}
\label{eq: mixingangles2}
N_{S_1}^{1,2} &\approx \mp \frac{1}{\sqrt{2}} ~,~ ~~~N_{S_2}^{1,2} \approx - \frac{1}{\sqrt{2}}
\\
N_{\nu_1}^{1,2} &\approx  \frac{v_d}{2 \left( M_D^2 - M_S^2 \right)} \big[ \left( y_{11} \pm y_{12} \right) M_D \pm \left( y_{21} \pm y_{22} \right) M_S \big]
\nonumber \\
N_{\nu_2}^{1,2} &\approx  \frac{v_d}{2 \left( M_D^2 - M_S^2 \right)} \big[ \left( y_{21} \pm y_{22} \right) M_D \pm \left( y_{11} \pm y_{12} \right) M_S \big]
\nonumber
~.
\end{align}
Furthermore, in this limit, the mass splitting, $\delta m_{\chi} \equiv m_{\chi_2} - m_{\chi_1}$, can be approximated as:
\begin{align}
\label{eq: splitting}
\delta m_{\chi} \approx \frac{v_d^2}{M_D^2-M_S^2} \, \Big|  &\left( y_{11} y_{12} + y_{21} y_{22} \right) M_D 
\nonumber \\
+ &\left( y_{11} y_{21} +  y_{12} y_{22} \right) M_S \Big| 
~.
\end{align}
From Eq.~\ref{eq: splitting} one can see that if either $y_{11} = y_{22} = 0$ or $y_{12} = y_{21} = 0$, then $\delta m_{\chi}= 0$. This can be understood from the symmetries of the Lagrangian of Eq.~\ref{eq: lagrangian1} as follows. The kinetic terms of $S_1$, $S_2$, $D_1$, $D_2$ possess a $U(1)^4$ symmetry. This is broken down to $U(1)_S \times U(1)_D$ by the bare masses $M_S$ and $M_D$ of Eq.~\ref{eq: lagrangian1}. At this point, $U(1)_S \times U(1)_D$ guarantees that the mass eigenstates of $M_0$ in Eq.~\ref{eq: massmatrix} will split into two separate degenerate pairs. Once the Yukawa couplings $y_{11}$, $y_{12}$, $y_{21}$, $y_{22}$ are turned on, both of these $U(1)$'s are further broken. However, in the limit that either $y_{11} = y_{22} = 0$ or $y_{12} = y_{21} = 0$ holds, then $U(1)_S \times U(1)_D$ is restored, and the mass spectrum once again decouples to two pairs of mass degenerate states. This argument holds to all orders in perturbation theory. 

To obtain a mass splitting as small as 3.55 keV, Yukawa couplings on the order of $y_{ij} \sim 10^{-2} \times (\tan \beta/50)(M_D/700 \,{\rm GeV})^{1/2}$ are generally required. As we will see in the next section, however, order one values for these quantities are necessary if $\chi_1 \chi_2$ coannihilations are to be efficient enough to obtain the desired thermal relic abundance without conflicting with constraints from gamma-ray observations of dwarf galaxies. This tension can be resolved if there is a cancellation in Eq.~\ref{eq: splitting}, such as arises for the choice $y_{11} \simeq y_{22} \simeq y_{21} \simeq -y_{12}$, with a small level of non-degeneracy needed for an $\mathcal{O}(\text{keV})$ mass splitting. This is technically natural and can arise from symmetries in the ultraviolet, which are expected to be broken at the two-loop level (for a similar approach, see Ref.~\cite{Falkowski:2014sma}).

\begin{figure}[t]
\includegraphics[width=3.5in]{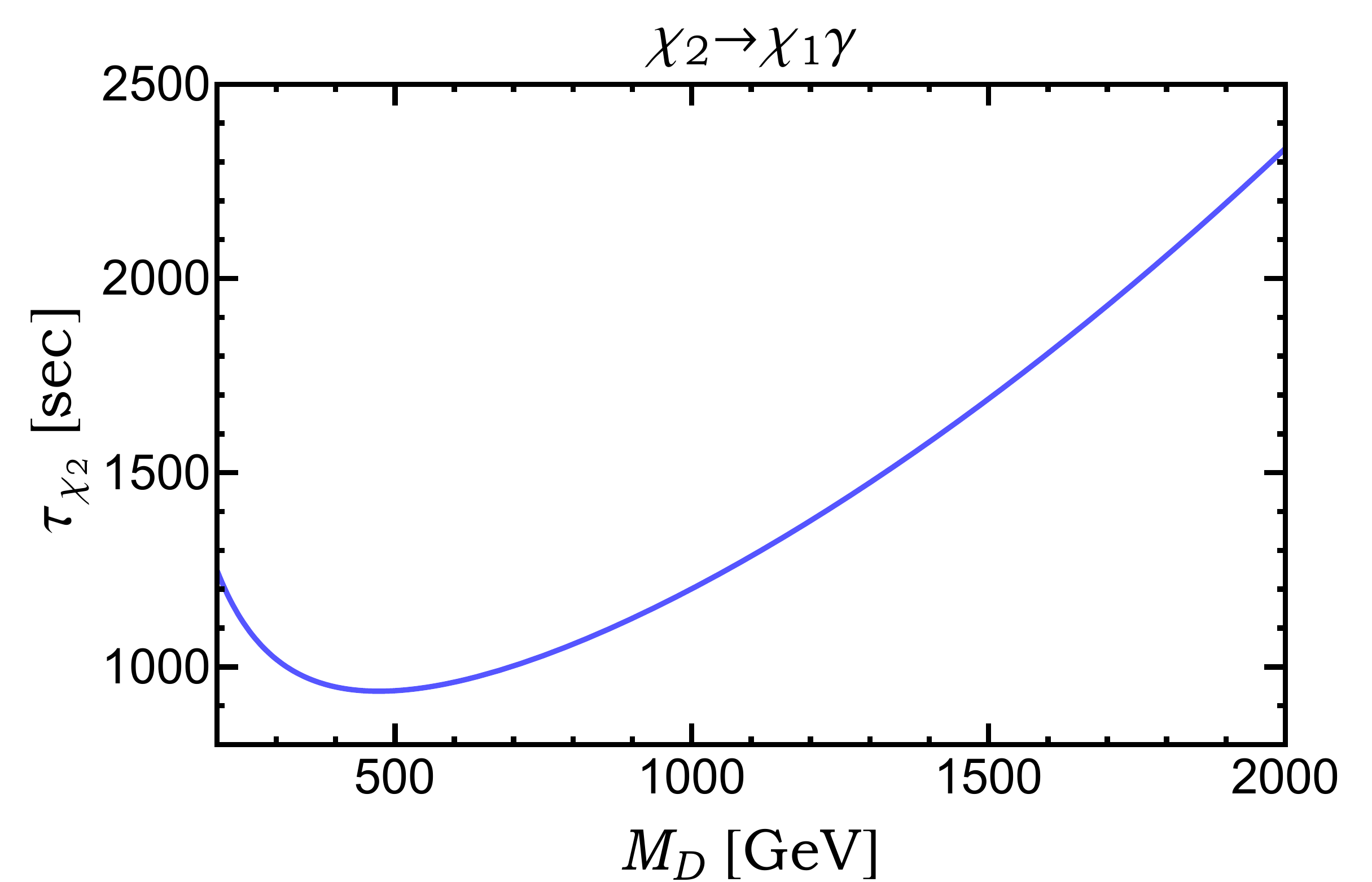} 
\caption{The lifetime for the decay $\chi_2 \rightarrow \chi_1 \gamma$, for $\delta m_{\chi}=3.55$ keV, $m_{\chi} \ll M_D, m_{H^\pm}$, $m_{H^{\pm}}=1$ TeV, $\tan{\beta} \gg 1$, $y_{11} \approx y_{21} \approx y_{22} \approx - y_{12} \approx 2.5$.}
\label{lifetimeplot}
\end{figure}

After electroweak symmetry breaking, $\chi_{1}$, $\chi_2$, and $\chi^{\pm}$ couple to the charged bosons $H^\pm$ and $W^\pm$, and a mass splitting, $\delta m_{\chi}$, is induced. As a result $\chi_2$ can decay via a radiative 2-body process $\chi_2 \to \chi_1 \gamma$ where $\chi^{\pm}$ and the charged bosons are exchanged in the loop (see Fig.~\ref{decaydiagram}). In the simple limit that $y_{11} = y_{22} = y_{21} = -y_{12} \equiv y$, $m_{H^\pm}, M_D \gg m_{\chi}$, and $\tan \beta \gg 1$, the $W^{\pm}$ loop contributes negligibly, and this decay width is given by:
\begin{equation}
\Gamma \approx \frac{e^2 y^4}{256 \pi^5} \frac{(\delta m_{\chi})^3  M^2_D}{(m^2_{H^\pm} -M^2_D)^2} \bigg[1-\frac{m^2_{H^\pm}}{m^2_{H^\pm} -M^2_D} \ln\bigg(\frac{m^2_{H^\pm}}{M^2_D}\bigg)\bigg]^2.
\end{equation}
A more general expression for this width is given in Appendix~\ref{analyticdecay}. We find that the most viable parameter space leads to lifetimes for $\chi_2$ that are on the order of an hour (see Fig.~\ref{lifetimeplot}). Furthermore, the 3-body decay to neutrinos, $\chi_2 \to \chi_1 \nu \nu$, remains kinematically open via an off-shell $Z$ boson. This later process, however, remains subdominant since it is suppressed by two additional powers of $\delta m_{\chi}$~\cite{Falkowski:2014sma}. 

In the region of parameter space under consideration, our model is not significantly constrained by observations of the cosmic microwave background (CMB) or of the light element abundances. In particular, for $\chi_2$ lifetimes less than $10^{12}$ s, the most stringent bounds from the CMB are on the $\chi_1 \chi_1$ annihilation cross section, and are weaker than the values considered in our study~\cite{planckdm}. Constraints from Big Bang Nucleosynthesis (BBN) are generally expressed in terms of the rescaled electromagnetic energy released, as a function of the $\chi_2$ lifetime. For lifetimes less than $10^6$ s, the upper bound is derived from the measured deuterium abundance~\cite{Feng:2003uy,Cyburt:2002uv}. Our model, however, predicts values for this quantity that are many orders of magnitude smaller than the existing upper limit.


\section{Results}
\label{results}

Due to the sizable number of free parameters in this model ($M_S$, $M_D$, $y_{11}$, $y_{22}$, $y_{12}$, $y_{21}$, $y_D$, $y_S$, $\alpha$, $\beta$, $m_{a_0}$, $m_{A_0}$, $B_a$), a wide range of phenomenology can emerge. We will simplify this to some extent by focusing on the parameters which yield $\delta m_{\chi} \simeq 3.55$ keV and $m_{\chi} \simeq 60$ GeV, which are within the range capable of generating both the 3.55 keV line and the Galactic Center gamma-ray excess~\cite{Calore:2014nla}.

\begin{figure}[t]
\includegraphics[width=3.5in]{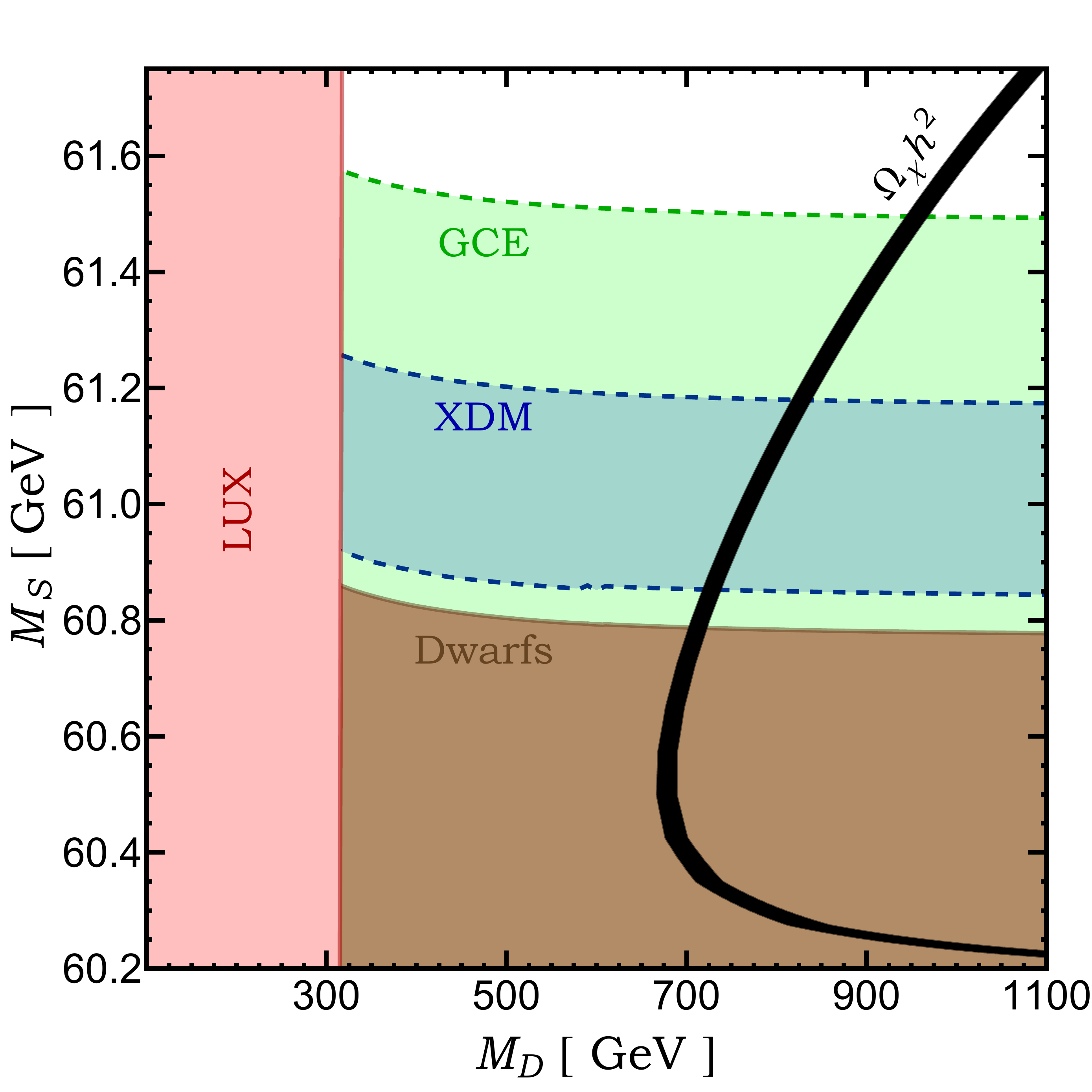} 
\caption{An example of the parameter space of our model, with $|\lambda^{a}_{11,22}|= 6$, $y_D=0$, $y_{11}=y_{21}=y_{22}=2.5$, $y_{12}\approx -2.5$, $\delta m_\chi = 3.55$ keV, $\theta = 3\times 10^{-5}$, $\tan \beta=50$, $m_a=120$ GeV, and $m_A=m_H=m_{H^\pm}=1$ TeV. The region yielding a thermal relic abundance equal to the cosmological dark matter density $\Omega_\chi h^2 = 0.1199 \pm 0.0027$ (solid black line) passes through the region that can generate the observed 3.55 keV signal (labelled ``XDM'') for values of $M_S  \approx 61$ GeV (about 1 GeV above the $a$ resonance). Also shown is the region capable of generating the Galactic Center gamma-ray excess (labelled ``GCE'', corresponding to $\sigma v =5\times 10^{-27}$ to $5\times 10^{-26}$ cm$^3$/s) and the regions that are excluded by gamma-ray observations of dwarf galaxies~\cite{fermidwarf} or by LUX~\cite{Akerib:2013tjd}.}
\label{parameterspace}
\end{figure}

In Fig.~\ref{parameterspace}, we show an example of a slice of the parameter space in this model. Here, we have adopted $|\lambda^{a}_{11,22}|= 6$, $y_D=0$, $y_{11}=y_{21}=y_{22}=2.5$, $y_{12}\approx -2.5$ (with $y_{12}$ fixed to give $\delta m_\chi=3.55$ keV throughout the $M_D-M_S$ plane shown), $\theta = 3\times 10^{-5}$, $\tan \beta=50$, $m_a=120$ GeV, and
$m_A=m_H=m_{H^{\pm}}=1$ TeV. The region yielding a thermal relic abundance equal to the cosmological dark matter density (solid black line) passes through the region that can generate the observed 3.55 keV line (labelled ``XDM'') for values of $M_S \approx 61$ GeV (about 1 GeV above the $a$ resonance). In particular, to provide an adequate fit to the 3.55 keV signal we demand that the up-scattering rate fits the measured fluxes within the $\delta \chi^2 < 3$ contour of Ref.~\cite{Cline:2014vsa} for promptly decaying excited states. Also shown is the region that is capable of generating the Galactic Center gamma-ray excess ($\sigma v =5\times 10^{-27}$ to $5\times 10^{-26}$ cm$^3$/s), as well as the regions that are excluded by gamma-ray observations of dwarf galaxies~\cite{fermidwarf} or by direct detection experiments~\cite{Akerib:2013tjd} (see Appendix~\ref{direct}). Constraints from the invisible width of the Higgs do not restrict any of the parameter space shown.

The process of $\chi_1 \chi_2$ coannihilation plays an important role in the determination of the thermal relic abundance in this region of parameter space.  In particular, in the limit at hand, $\lambda^h_{12}/\lambda^h_{11} \sim 2 M_D/M_S \gg 1$, large coannihilation rates are possible without large elastic scattering cross sections with nuclei (for details, see Appendices~\ref{appendixa} and~\ref{direct}). In addition, resonant annihilation and coannihilation through the $a$ and the SM-like Higgs significantly deplete the thermal abundance for $M_S\simeq 60$ GeV and 62.5 GeV, respectively. The relic abundance is relatively insensitive to the parameters scanned over in Fig.~\ref{parameterspace}, and is within approximately an order of magnitude of the measured quantity over the entire plane shown. Lastly, we note that up-scattering and decay rates are insensitive to the parameter $\theta$. If we had chosen to set this quantity to zero (decoupling $a$ from the pseudoscalar of the 2HDM), we can still generate the 3.55 keV line, but without a mechanism to produce the Galactic Center gamma-ray excess and without constraints from gamma-ray observations of dwarf galaxies.

\section{Summary and Conclusions}
\label{conclusion}

It has been previously proposed that dark matter scattering into an excited state (eXciting Dark Matter, or XDM) could be responsible for the 3.55 keV line observed from Galaxy Clusters without conflicting with the lack of such a signal from dwarf galaxies~\cite{Finkbeiner:2014sja}. Such a model could also potentially generate Fermi's gamma-ray excess from the Galactic Center. Most of the XDM model building discussed in the literature has focused on scenarios in which the dark matter interacts through a light mediator, with no significant couplings between the dark sector and the Standard Model. Here, instead of hidden sector, we have considered a model in which the dark matter directly annihilates into Standard Model fermions through the near resonant exchange of a pseudoscalar, $a$, which also efficiently mediates the process of up-scattering, $\chi_1 \chi_1 \rightarrow \chi_2 \chi_2$. This pseudoscalar is a mixture of a Standard Model singlet and the pseudoscalar appearing from a two-Higgs doublet model. The dark matter itself is a mixture of two Standard Model gauge singlets and the neutral components of two $SU(2)_W$ doublets. This allows us to generate a 3.55 keV mass splitting between the two lightest mass eigenstates, and enables for the rapid decay of $\chi_2 \rightarrow \chi_1 \gamma$.

We have identified regions of parameter space in our model that can simultaneously generate the 3.55 keV line and the Galactic Center gamma-ray excess, while remaining consistent with all constraints from colliders, direct detection experiments, and gamma-ray observations of dwarf galaxies. 
Coannihilations between $\chi_1$ and $\chi_2$ can play an important role in determining the thermal relic abundance of dark matter in this model.

\bigskip \bigskip \bigskip

{\it Acknowledgments}: AB is supported by the Kavli Institute for Cosmological Physics at the University of Chicago through grant NSF PHY-1125897. 
AD is supported by a Fermilab Fellowship in Theoretical Physics. DH is supported by the US Department of Energy under contract DE-FG02-13ER41958. Fermilab is operated by Fermi Research Alliance, LLC, under Contract No. DE-AC02-07CH11359 with the US Department of Energy.


\pagebreak

\appendix
\onecolumngrid
\section{Higgs and Gauge Couplings}
\label{appendixa}

In this appendix, we provide analytic forms for the couplings of $\chi_1 \chi_1$, $\chi_2 \chi_2$, and $\chi_1 \chi_2$ to the light Higgs bosons ($a$, $h$) and to the $Z$.  These couplings are defined according to the following terms in the Lagrangian:
\begin{eqnarray}
\label{eq:OriginalLagrangian}
\mathcal{L} &\supset& \lambda^a_{11} a  \bar{\chi}_1 i \gamma^5 \chi_1 + \lambda^a_{22} a  \bar{\chi}_2 i \gamma^5 \chi_2 + \lambda^a_{12} a  \bar{\chi}_1 i \gamma^5 \chi_2  + \lambda^h_{11} h  \bar{\chi}_1  \chi_1 + \lambda^h_{22} h  \bar{\chi}_2  \chi_2 + \lambda^h_{12} h  \bar{\chi}_1  \chi_2 \nonumber \\
&+& g_{11} Z_{\mu} \bar{\chi}_1 \gamma^{\mu} \gamma^5 \chi_1 + g_{22} Z_{\mu} \bar{\chi}_2 \gamma^{\mu} \gamma^5 \chi_2 + g_{12} Z_{\mu} \bar{\chi}_1 \gamma^{\mu} \gamma^5 \chi_2. 
\end{eqnarray}
As discussed in the text, the field $\chi_2$ requires a transformation of the form $\chi_2 \rightarrow i \gamma^5 \chi_2$ in order to ensure a positive value for its mass term. After this field redefinition, the above Lagrangian appears as follows:
\begin{eqnarray}
\mathcal{L} &\supset& \lambda^a_{11} a  \bar{\chi}_1 i \gamma^5 \chi_1 - \lambda^a_{22} a  \bar{\chi}_2 i \gamma^5 \chi_2 - \lambda^a_{12} a  \bar{\chi}_1  \chi_2  + \lambda^h_{11} h  \bar{\chi}_1  \chi_1 - \lambda^h_{22} h  \bar{\chi}_2  \chi_2 + \lambda^h_{12} h  \bar{\chi}_1 i \gamma^5 \chi_2 \nonumber \\
&+& g_{11} Z_{\mu} \bar{\chi}_1 \gamma^{\mu} \gamma^5 \chi_1 + g_{22} Z_{\mu} \bar{\chi}_2 \gamma^{\mu} \gamma^5 \chi_2 + i g_{12} Z_{\mu} \bar{\chi}_1 \gamma^{\mu} \chi_2. 
\end{eqnarray}

The couplings are given by:
\begin{eqnarray}
\lambda^a_{11} &=& \cos \theta \, [y_S N^1_{S_1} N^1_{S_2} + y_D N^1_{\nu_1} N^1_{\nu_2}] + \frac{\sin \theta \sin \beta}{\sqrt{2}} \, [y_{11} N^1_{S_1} N^1_{\nu_1} + y_{21} N^1_{S_2} N^1_{\nu_1} - y_{22} N^1_{S_2} N^1_{\nu_2} - y_{12} N^1_{S_1} N^1_{\nu_2}]  \nonumber \\
\lambda^a_{22} &=& \cos \theta \, [y_S N^2_{S_1} N^2_{S_2} + y_D N^2_{\nu_1} N^2_{\nu_2}] + \frac{\sin \theta \sin \beta}{\sqrt{2}} \, [y_{11} N^2_{S_1} N^2_{\nu_1} + y_{21} N^2_{S_2} N^2_{\nu_1} - y_{22} N^2_{S_2} N^2_{\nu_2} - y_{12} N^2_{S_1} N^2_{\nu_2}] \nonumber \\
\lambda^a_{12} &=& \cos \theta \, [y_S (N^1_{S_1} N^2_{S_2}+N^2_{S_1}N^1_{S_2})+ y_D (N^1_{\nu_1} N^2_{\nu_2}+ N^2_{\nu_1}N^1_{\nu_2})]
+ \frac{\sin \theta \sin \beta}{\sqrt{2}} \, [y_{11} (N^1_{S_1} N^2_{\nu_1}+N^2_{S_1} N^1_{\nu_1}) \nonumber \\
&+& y_{21} (N^1_{S_2} N^2_{\nu_1} 
+N^2_{S_2} N^1_{\nu_1}) - y_{22} (N^1_{S_2} N^2_{\nu_2}+N^2_{S_2} N^1_{\nu_2}) - y_{12} (N^1_{S_1} N^2_{\nu_2}+N^2_{S_1} N^1_{\nu_2})]  \nonumber \\
\lambda^h_{11} &=& -\frac{\cos \beta}{\sqrt{2}} \bigg[y_{11} N^1_{S_1} N^1_{\nu_1} +  y_{21} N^1_{S_2} N^1_{\nu_1}+ y_{22} N^1_{S_2} N^1_{\nu_2} +  y_{12} N^1_{S_1} N^1_{\nu_2} \bigg] \nonumber \\
\lambda^h_{22} &=& -\frac{\cos \beta}{\sqrt{2}} \bigg[y_{11} N^2_{S_1} N^2_{\nu_1} +  y_{21} N^2_{S_2} N^2_{\nu_1}+ y_{22} N^2_{S_2} N^2_{\nu_2} +  y_{12} N^2_{S_1} N^2_{\nu_2} \bigg] \nonumber \\
\lambda^h_{12} &=& -\frac{\cos \beta}{\sqrt{2}} \bigg[y_{11} (N^1_{S_1} N^2_{\nu_1} +N^2_{S_1} N^1_{\nu_1}) +  y_{21} (N^1_{S_2} N^2_{\nu_1} +N^2_{S_2} N^1_{\nu_1}) +  y_{22} (N^1_{S_2} N^2_{\nu_2} +N^2_{S_2} N^1_{\nu_2}) +  y_{12} (N^1_{S_1} N^2_{\nu_2} +N^2_{S_1} N^1_{\nu_2})  \bigg] \nonumber \\
g_{11} &= & -\frac{g}{4 c_W} \bigg[ (N^1_{\nu_1})^2-(N^1_{\nu_2})^2\bigg] \nonumber \\
g_{22} &=& -\frac{g}{4 c_W} \bigg[ (N^2_{\nu_1})^2-(N^2_{\nu_2})^2\bigg] \nonumber \\
g_{12} &=& -\frac{g}{2 c_W} \bigg[ N^1_{\nu_1} N^2_{\nu_1}- N^1_{\nu_2} N^2_{\nu_2} \bigg] \nonumber 
,
\end{eqnarray}
where the mixing angles are defined in Eq.~\ref{eq: mixingangles1}, and $g$ and $c_W$ are the $SU(2)_W$ coupling constant and cosine of the Weinberg angle, respectively. Note that in the limit of small $a_0$-$A_0$ mixing, $\lambda_{11, 22}^a  \approx  \pm y_S  \cos \theta/2$.

\section{Decay}
\label{analyticdecay}

In this appendix, we provide formulae describing the decay $\chi_2 \rightarrow \chi_1 \gamma$. We note that a convenient gauge choice is \emph{non-linear R gauge} due to the vanishing of the $\gamma W^+ G^-$ vertex (see e.g. Ref.~\cite{Haber:1988px}). The width for this process is given by:
\begin{equation}
\Gamma(\chi_2 \rightarrow \chi_1\gamma)=\frac{g^2_{\rm eff}}{8\pi} \bigg(\frac{\Delta}{m_{\chi_2}}\bigg)^3,
\end{equation}
where $\Delta \equiv m^2_{\chi_2} - m^2_{\chi_1}$. The effective coupling in this expression is given by:
\begin{eqnarray}
g_{\rm eff} &=& \frac{e \epsilon_1}{8 \pi^2} \bigg[   -(\epsilon_1 \epsilon_2 g_{1L} g_{2L}-g_{1R} g_{2R}) [\epsilon_2 m_{\chi_2} (I_{g2}-J_g-K_g) +\epsilon_1 m_{\chi_1} (J_g-K_g)] -2M_D (\epsilon_1 g_{1L}g_{2R}-\epsilon_2g_{1R}g_{2L}) J_g \nonumber \\
&+&\frac{1}{4} (\lambda_{1L} \lambda_{2L} -\epsilon_1 \epsilon_2 \lambda_{1R} \lambda_{2R}) [\epsilon_2 m_{\chi_2} (I_{s2} -K_s)-\epsilon_1 m_{\chi_1} K_s] + \frac{1}{4} M_D (\epsilon_2 \lambda_{1L} \lambda_{2R} -\epsilon_1 \lambda_{1R} \lambda_{2L})I_s  \nonumber \\
&+&\frac{1}{4} (\lambda^G_{1L} \lambda^G_{2L} - \epsilon_1 \epsilon_2 \lambda^G_{1R} \lambda^G_{2R})[\epsilon_2 m_{\chi_2} (I_{g2}-K_{g}) - \epsilon_1 m_{\chi_1} K_{g}] + \frac{1}{4} M_D(\epsilon_2 \lambda^G_{1L}\lambda^G_{2R}-\epsilon_1 \lambda^G_{1R}\lambda^G_{2L}) I_{g}  \bigg],
\label{geff}
\end{eqnarray}
where $\epsilon_1$ and $\epsilon_2$ are the signs of the first two eigenvalues of the mass matrix, $M_0$, as given in Eq.~\ref{eq: massmatrix}. The couplings in the above expression are given by:
\begin{equation}
g_{1L} \equiv -\frac{g}{\sqrt{2}} N^1_{\nu_2},  \,\,\,\,\,\, \,\,\,\,
g_{1R} \equiv -\frac{g}{\sqrt{2}} N^1_{\nu_1},\,\,\,\,\,\, \,\,\,\,
g_{2L} \equiv -\frac{g}{\sqrt{2}} N^2_{\nu_2},\,\,\,\,\,\, \,\,\,\,
g_{2R} \equiv -\frac{g}{\sqrt{2}} N^2_{\nu_1},
\end{equation}
\begin{equation}
\lambda_{1L} \equiv  -\sin \beta \, (y_{22} N^1_{S_2} + y_{12} N^1_{S_1}) ,  \,\,\,\,\,\, \,\,\,\, \nonumber
\lambda_{2L} \equiv -\sin \beta \, (y_{22} N^2_{S_2} + y_{12} N^2_{S_1}),
\end{equation}
\begin{equation}
\lambda_{1R} \equiv -\sin \beta \, (y_{11} N^1_{S_1}+ y_{21} N^1_{S_2}),  \,\,\,\,\,\,  \,\,\,\,\nonumber
\lambda_{2R} \equiv -\sin \beta \, (y_{11} N^2_{S_1} + y_{21} N^2_{S_2}),  
\end{equation}
\begin{equation}
\lambda^G_{1L} \equiv \cos \beta \, (y_{22} N^1_{S_2} + y_{12} N^1_{S_1}) ,  \,\,\,\,\,\, \,\,\,\, \nonumber
\lambda^G_{2L} \equiv \cos \beta \, (y_{22} N^2_{S_2} + y_{12} N^2_{S_1}) ,
\end{equation}
\begin{equation}
\lambda^G_{1R} \equiv \cos \beta \, (y_{11} N^1_{S_1} + y_{21} N^1_{S_2}),  \,\,\,\,\,\,  \,\,\,\,\nonumber
\lambda^G_{2R} \equiv \cos \beta \, (y_{11} N^2_{S_1} + y_{21} N^2_{S_2}).
\end{equation}

\twocolumngrid

Lastly, Eq.~\ref{geff} contains a number of integrals, defined as follows:
\begin{eqnarray}
I_{g} &\equiv& \frac{1}{\Delta} \int^1_0 \frac{dx}{1-x} \log X_g, \\
I_{g2} &\equiv& \frac{1}{\Delta} \int^1_0 dx \log X_g, \nonumber \\
J_{g} &\equiv& \frac{1}{\Delta} \int^1_0 \frac{dx}{1-x} \log X'_g ,\nonumber \\
I_{s} &\equiv& \frac{1}{\Delta} \int^1_0 \frac{dx}{1-x} \log X_s, \nonumber\\
I_{s2} &\equiv& \frac{1}{\Delta} \int^1_0 dx \log X_s, \nonumber\\
J_{s} &\equiv& \frac{1}{\Delta} \int_0^1 \frac{dx}{1-x} \log{X_s^\prime} , \nonumber \\
K_g &\equiv &-\frac{1}{\Delta} (1 +M^2_D I_g +m^2_W J_g -m^2_{\chi_2} I_{g2}) , \nonumber\\
K_s &\equiv& -\frac{1}{\Delta} (1 +M^2_D I_s +m^2_{H^\pm} J_s -m^2_{\chi_2} I_{s2}) , \nonumber
\end{eqnarray}
where
\begin{eqnarray}
X_g \equiv \frac{M^2_D x +m^2_W (1-x)-m^2_{\chi_2}x(1-x)}{M^2_D x +m^2_W (1-x)-m^2_{\chi_1}x(1-x)},\\
X'_g \equiv \frac{m^2_W x +M^2_D (1-x)-m^2_{\chi_2}x(1-x)}{m^2_W x +M^2_D (1-x)-m^2_{\chi_1}x(1-x)},\nonumber \\
X_s \equiv \frac{M^2_D x +m^2_{H^\pm} (1-x)-m^2_{\chi_2}x(1-x)}{M^2_D x +m^2_{H^\pm} (1-x)-m^2_{\chi_1}x(1-x)}\nonumber \\
X_s^\prime \equiv \frac{m_{H^\pm}^2 x + M_D^2 (1-x) - m_{\chi_2}^2 x (1-x)}{m_{H^\pm}^2 x + M_D^2 (1-x) - m_{\chi_1}^2 x (1-x)}. \nonumber
\end{eqnarray}

\section{Direct Detection}
\label{direct}

The elastic scattering cross section between the dark matter, $\chi_1$, and a nucleus with atomic number $Z$ and atomic mass $A$ is given by:
\begin{eqnarray}
\sigma^{\rm elastic}_{0} = \frac{4 \mu^2_{\chi, N}}{\pi} \bigg[Z f_p + (A-Z) f_n\bigg]^2,
\end{eqnarray}
where $\mu_{\chi, N}$ is the reduced mass of the system and the nucleon level couplings are given by:
\begin{equation}
f_{p,n} = m_{p,n} \bigg[ \sum_{q=u,d,s} \frac{a_q}{m_q} f^{(p,n)}_{T_q} + \frac{2}{27} f^{(p,n)}_{TG} \sum_{q=c,b,t}  \frac{a_q}{m_q}\bigg],
\end{equation}
and
\begin{eqnarray}
\frac{a_q}{m_q} = \frac{1}{v} \bigg[-\frac{\lambda^h_{11}}{m^2_h} +\frac{\lambda^H_{11} q_{\beta}}{m^2_H} \bigg],
\end{eqnarray}
where $q_{\beta}=\cot \beta$ ($-\tan \beta$) for up-type (down-type) quarks. 

In addition, direct detection experiments can also detect inelastic events, $\chi_1 N \rightarrow \chi_2 N$.  For $\delta m_{\chi} \lsim v^2 \mu_{\chi,N}/2$, the dark matter particles typically possess enough kinetic energy to scatter into the excited state. The cross section for inelastic scattering is given by:
\begin{eqnarray}
\sigma^{\rm inelastic}_{0} = \frac{\mu^2_{\chi, N}}{\pi} \, F \,\bigg[Z f_p + (A-Z) f_n\bigg]^2,
\end{eqnarray}
where in this case, 
\begin{eqnarray}
f_p=\frac{g \sin \theta_W g_{12}}{4 m^2_Z} \left( \cot{\theta_W} - 3 \tan{\theta_W} \right) \\
f_n=-\frac{g \sin \theta_W g_{12}}{4 m^2_Z} \left( \cot{\theta_W} + \tan{\theta_W} \right) , \nonumber
\end{eqnarray}
and the following factor accounts for the kinematic suppression associated with inelastic scattering:
\begin{equation}
F=\bigg[ \frac{s^2-2(m^2_{\chi_2} + m^2_N)s + (m^2_{\chi_2}-m^2_{N})^2}{s^2-2(m^2_{\chi_1} + m^2_N)s + (m^2_{\chi_1}-m^2_{N})^2}\bigg]^{1/2},
\end{equation}
and
\begin{equation}
\sqrt{s} \approx m_N \bigg[1+\frac{1}{2} \bigg(\frac{\mu_{\chi, N} v}{m_N} \bigg)^2\bigg] + m_{\chi_1} \bigg[1+\frac{1}{2} \bigg(\frac{\mu_{\chi, N} v}{m_{\chi_1}} \bigg)^2\bigg],
\end{equation}
where $v \sim 10^{-3}$. Note that $F \rightarrow 0$ in the limit of $v \rightarrow v_{\rm min} = \sqrt{2 \delta m_{\chi}/\mu_{\chi, N}}$, below which inelastic scattering is not possible. In contrast, for the mass splitting and masses under consideration in this paper (and for typical dark matter velocities in the local Milky Way), $v \gg v_{\rm min}$ and $F \sim 1$.

\bibliography{keVGC}

\end{document}